\documentclass{bmvc2k}
\usepackage{hyperref}
\usepackage{graphicx}
\usepackage{amssymb}
\usepackage{adjustbox}
\usepackage{colortbl}
\usepackage{comment}
\usepackage{multirow}

\newcommand{\ttt}{\boldsymbol{\theta}}
\newcommand{\ppp}{\boldsymbol{\phi}}

\newcommand{\xx}{\mathbf{x}}
\newcommand{\aaa}{\mathbf{a}}
\newcommand{\mm}{\mathbf{m}}

\newcommand{\zz}{\mathbf{z}}
\newcommand{\real}{\mathbb{R}}

\definecolor{gray}{cmyk}{0.86,0.86,0.86,0.86}

\title{Looking at the whole picture: constrained unsupervised anomaly segmentation}
\addauthor{Julio Silva-Rodríguez}{jjsilva@upv.es}{1}
\addauthor{Valery Naranjo}{vnaranjo@dcom.upv.es}{2}
\addauthor{Jose Dolz}{jose.dolz@etsmtl.ca}{3}

\addinstitution{
 Institute of Transport and Territory\\
 Universitat Politècnica de València\\
 Valencia, Spain
}
\addinstitution{
 Institute of Research and Innovation in Bioengineering\\
 Universitat Politècnica de València\\
 Valencia, Spain
}
\addinstitution{
 LIVIA Laboratory\\
 École de Technologie Supérieure (ETS)\\
 Montreal, Canada
}

\runninghead{Silva-Rodríguez et al}{Looking at the whole picture}

\begin{document}

\maketitle

\begin{abstract}

Current unsupervised anomaly localization approaches rely on generative models to learn the distribution of normal images, which is later used to identify potential anomalous regions derived from errors on the reconstructed images. However, a main limitation of nearly all prior literature is the need of employing anomalous images to set a class-specific threshold to locate the anomalies. This limits their usability in realistic scenarios, where only normal data is typically accessible. Despite this major drawback, only a handful of works have addressed this limitation, by integrating supervision on attention maps during training. In this work, we propose a novel formulation that does not require accessing images with abnormalities to define the threshold. Furthermore, and in contrast to very recent work, the proposed constraint is formulated in a more principled manner, leveraging well-known knowledge in constrained optimization. In particular, the equality constraint on the attention maps in prior work is replaced by an inequality constraint, which allows more flexibility. In addition, to address the limitations of penalty-based functions we employ an extension of the popular log-barrier methods to handle the constraint. Comprehensive experiments on the popular BRATS'19 dataset demonstrate that the proposed approach substantially outperforms relevant literature, establishing new state-of-the-art results for unsupervised lesion segmentation. 

\end{abstract}

%-------------------------------------------------------------------------
\section{Introduction}
\label{sec:intro}

Under the supervised learning paradigm, deep learning models have achieved astonishing performance in a wide range of applications. Nevertheless, a main limitation of these models is the large amount of labeled data required for training. Obtaining such curated labeled datasets is a cumbersome process prone to subjectivity, which makes access to sufficient training data difficult in practice. This problem is further magnified in the context of medical image segmentation, where labeling involves assigning a category to each image pixel 
resulting impractical when volumetric data is involved. 
In addition, even if annotated images are available, there exist some applications, such as brain lesion detection, where large intraclass variations are not captured during training, failing to cover the broad range of abnormalities that might be present in a scan. This makes that, in a fully-supervised setting, deep models might have difficulties when learning from such class-imbalanced training sets. Thus, considering the scarcity and the diversity of target objects in these scenarios, lesion segmentation is typically modeled as an anomaly localization task, which is trained in an unsupervised manner. In particular, the training dataset contains only \textit{normal} images and \textit{abnormal} images are not accessible during training. 

A common strategy for unsupervised anomaly segmentation is to model the distribution of normal images, for which generative models, such as generative adversarial networks (GANs) \cite{schlegl2019f,schlegl2017unsupervised,andermatt2018pathology,ravanbakhsh2019training,baur2020steganomaly,sun2020adversarial} and variational auto-encoders (VAEs) \cite{chen2018unsupervised,pawlowski2018unsupervised,sabokrou2018avid,Chen2020UnsupervisedPriorMedIA,zimmerer2019context} have been widely employed. To achieve this, input images are compared to their reconstructed normal %\textit{healthy} 
counterparts, which are recovered from the learned distribution, and anomalies are identified from the reconstruction error. Nevertheless, these methods require to learn a threshold to estimate the pixel-wise difference between the input and its reconstructed image in order to localize abnormalities. As the threshold needs to be computed based on abnormal training images, this limits their usability in realistic scenarios, where only normal data is provided. Inspired by the observations that attention-based supervision can alleviate the need of large labeled training data \cite{li2018tell}, class-activation maps have been integrated in the training. In particular, \cite{venkataramanan2020attention} leverage the generated attention maps as an additional supervision cue, enforcing the network to provide attentive regions covering the whole context in normal images. This term was formulated as an equality constraint with the form of a L$_1$ penalty over each individual pixel. Nevertheless, we found that explicitly forcing the network to produce maximum attention values across each pixel does not achieve satisfactory results in the context of brain lesion segmentation. In addition, recent literature in constrained optimization for deep neural networks suggests that simple penalties --such as the function used in \cite{venkataramanan2020attention}-- might not be the optimal solution to constraint the output of a CNN \cite{kervadec2019constrained}.

Based on these observations, we propose a novel formulation for unsupervised semantic segmentation of brain lesions in medical images. The key contributions of our work can be summarized as follows:

\begin{itemize}
    \item A novel constrained formulation for unsupervised %pixel-wise 
    anomaly localization, which integrates an auxiliary size-constrained loss to force the network to generate class activation masks (CAMs) that cover the whole context in normal images.
    
    \item In particular, size information is imposed through inequality constraints on the region proportion of generated CAMs, which give more flexibility than the pixel-wise equality constraint in \cite{venkataramanan2020attention}. In addition, to address the limitations of penalty-based functions, we resort to an extended version of the standard log-barrier.
    
    \item Furthermore, while our method yields significant improvements when anomalous images are used to define a class-specific threshold to locate anomalies --following the literature--, our formulation still outperforms existing approaches without accessing to anomalous images, which contrasts to most prior works.
    
    \item We benchmark the proposed model against a relevant body of literature on the popular BRATS challenge dataset. Comprehensive experiments demonstrate the superior performance of our model, % compared to prior works, 
    establishing a new state-of-the-art for this task.
\end{itemize}

\section{Related Work}
\label{sec:rw}

\paragraph{Unsupervised Anomaly Segmentation.}Unsupervised anomaly segmentation aims at identifying abnormal pixels on test images, containing, for example, lesions on medical images \cite{baur2020steganomaly,chen2018unsupervised}, defects in industrial images \cite{bergmann2018improving,liu2020towards,venkataramanan2020attention} or abnormal events in videos \cite{abati2019latent,ravanbakhsh2019training}. A main body of the literature has explored unsupervised deep (generative) representation learning to learn the distribution from normal data. The underlying assumption is that a model trained on normal data will not be able to reconstruct anomalous regions, and the reconstructed difference can therefore be used as an anomaly score. Under this learning paradigm, generative adversarial networks (GAN) \cite{goodfellow2014generative} and variational auto-encoders (VAE) \cite{kingma2013auto} are typically employed. Nevertheless, even though GAN and VAE model the latent variable, the manner in which they approximate the distribution of a set of samples differs. GAN-based approaches \cite{schlegl2019f,schlegl2017unsupervised,andermatt2018pathology,ravanbakhsh2019training,baur2020steganomaly,sun2020adversarial} approximate the distribution by optimizing a generator to map random samples from a prior distribution in the latent space into data points that a trained discriminator cannot distinguish. On the other hand, data distribution is approximated in VAE by using variational inference, where an encoder approximates the posterior distribution in the latent space and a decoder models the likelihood \cite{chen2018unsupervised,pawlowski2018unsupervised,sabokrou2018avid,Chen2020UnsupervisedPriorMedIA,zimmerer2019context,dehaene2019iterative}. In the context of medical images, and more relevant to this work, several works have proposed different improvements on VAEs \cite{chen2018unsupervised,pawlowski2018unsupervised,Chen2020UnsupervisedPriorMedIA}, to overcome specific limitations. For example, to handle the lack of consistency in the learned latent representation on prior works, \cite{chen2018unsupervised} included a constraint that helps mapping an image containing abnormal anatomy close to its corresponding healthy image in the latent space. A detailed survey on unsupervised anomaly localization in medical imaging can be found in \cite{Baur2021AutoencodersStudy}. Nevertheless, a main limitation of these approaches is that the threshold to estimate the pixel-wise anomaly score has to be computed in images with anomalies, which might not be available in practice. To alleviate this issue, 
\cite{venkataramanan2020attention} propose to integrate attention maps from Grad-CAM \cite{Selvaraju2020Grad-CAM:Localization} during the training as supervisory signals. In particular, in addition to standard learning objectives, authors employ an auxiliary loss that tries to maximize the attention maps on normal images by including an equality constraint with the form of a L$_1$ penalty over each individual pixel.

\vspace{-4mm}

\paragraph{Constrained segmentation.}Imposing global constraints on the output predictions of deep CNNs has gained attention recently, particularly in weakly supervised segmentation. These constraints can be embedded into the network outputs in the form of direct loss functions, which guide the network training when fully labeled images are not accessible. For example, a popular scenario is to enforce the softmax predictions to satisfy a prior knowledge on the size of the target region. Jia et \textit{al.} \cite{jia2017constrained} employed a L$_2$ penalty to impose equality constraints on the size of the target regions in the context of histopathology image segmentation. In \cite{zhang2017curriculum}, authors leverage the target properties by enforcing the label distribution of predicted images to match an inferred label distribution of a given image, which is achieved with a KL-divergence term. Similarly, Zhou et \textit{al.} \cite{zhou2019prior} proposed a novel loss objective in the context of partially labeled images, which integrated an auxiliary term, based on a KL-divergence, to enforce that the average output size distributions of different organs approximates their empirical distributions, obtained from fully-labeled images. While the equality-constrained formulations proposed in these works are very interesting, they assume exact knowledge of the target size prior. In contrast, inequality constraints can relax this assumption, allowing much more flexibility. In \cite{pathak2015constrained}, authors imposed inequality constraints on a latent distribution --which represents a “fake” ground truth-- instead of the network output, to avoid the computational complexity of directly using Lagrangian-dual optimization. Then, the network parameters are optimized to minimize the KL divergence between the network softmax probabilities and the latent distribution. Nevertheless, their formulation is limited to linear constraints. More recently, inequality constraints have been tackled by augmenting the learning objective with a penalty-based function, e.g., L$_2$ penalty, which can be imposed within a continuous optimization framework \cite{kervadec2019constrained,kervadec2019curriculum,bateson2019constrained}, or in the discrete domain \cite{peng2020discretely}. Despite these methods have demonstrated remarkable performance in weakly supervised segmentation, they require that prior knowledge, \textit{exact} or \textit{approximate}, is given. This contrasts with the proposed approach, which is trained on data without anomalies, and hence the size of the target is zero.

\vspace{-5mm}

\section{Methods}
\label{sec:methods}

\vspace{-1mm}

\begin{figure*}[h!]
\begin{center}
\includegraphics[width=1\textwidth]{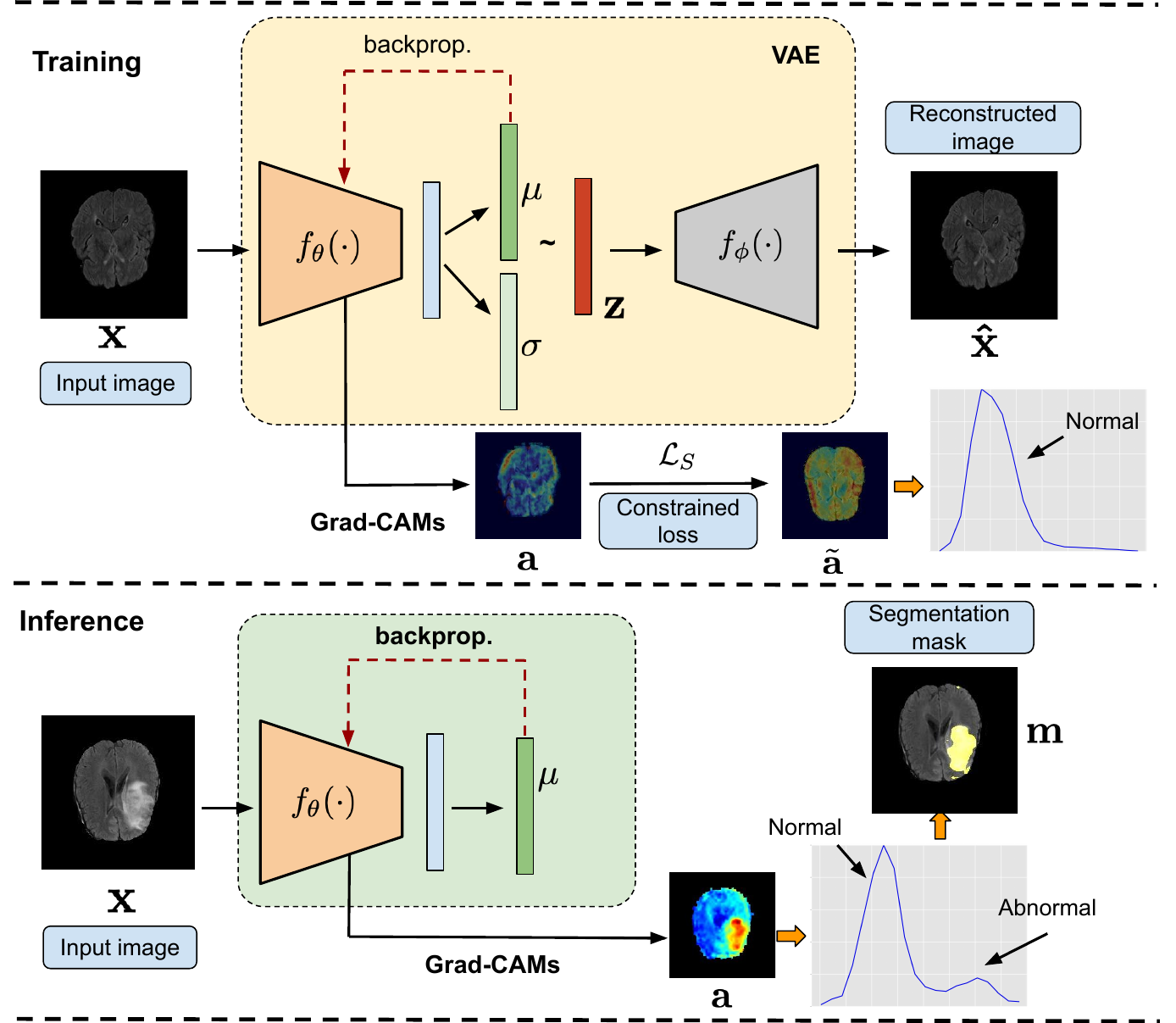}
\caption{\textbf{Method overview}. Following the standard literature, the VAE is optimized to maximize the evidence lower bound (ELBO), which satisfies Eq. \ref{eq:VAE_vanilla}. In addition, we include a size regularizer (in the form of a size-constrained loss $\mathcal{L}_s$) on the attention maps $\aaa$, to force the network to search in the whole image. 
At inference, the attention map is thresholded to obtain the final segmentation mask $\mm$.}
\label{fig:summary}
\end{center}
\vspace{-4mm}
\end{figure*}

An overview of our method is presented in Fig. \ref{fig:summary}, and we describe each component below.

\vspace{-1mm}

\vspace{-4mm}

\paragraph{Preliminaries.}Let us denote the set of unlabeled training images as $\mathcal{D} = \{\xx_n\}_{n=1}^N$, where $\xx_i \in \mathcal{X} \subset \real^{\Omega_i}$ represents the \textit{i}$^{th}$ image and $\Omega_i$ denotes the spatial image domain. This dataset contains only normal images, e.g., healthy images in the medical context. We also define an encoder, $f_{\ttt}(\cdot) : \mathcal{X} \rightarrow \mathcal{Z}$, parameterized by $\ttt$, which is optimized to project normal data points in $\mathcal{D}$ into a manifold represented by a lower dimensionality $d$, $\zz \in \mathcal{Z} \subset \real^{d}$. Furthermore, a decoder $f_{\ppp}(\cdot) : \mathcal{Z} \rightarrow \mathcal{X}$ parameterized by $\ppp$ tries to reconstruct an input image $\xx \in \mathcal{X}$ from $\zz \in \mathcal{Z}$, which results in $\hat \xx=f_{\ppp}(f_{\ttt}(\xx))$.

\subsection{Vanilla VAE}

A Variational Autoencoder (VAE) is an encoder-decoder style generative model, which is currently the dominant strategy for unsupervised anomaly location. Training a VAE consists on minimizing a two-term loss function, which is equivalent to maximize the evidence lower-bound (ELBO) \cite{kingma2013auto}:

\vspace{-2mm}

\begin{equation}
\mathcal{L}_{VAE} = \mathcal{L}_{R}(\xx, \hat{\xx}) + \beta\mathcal{L}_{KL}(q_{\ttt}(\zz|\xx) || p(\zz))
\label{eq:VAE_vanilla}
\end{equation}

\noindent where $\mathcal{L}_{R}$ is the reconstruction error term between the input and its reconstructed counterpart. The right-hand term is the Kullback-Leibler (KL) divergence (weighted by $\beta$) between the approximate posterior $q_{\ttt}(\zz|\xx)$ and the prior $p(\zz)$, which acts as a regularizer, penalizing approximations for $q_{\ttt}(\zz|\xx)$ that differ from the prior.

\vspace{-2mm}

\subsection{Size regularizer via VAE attention}
\label{ssec:sizereg}

Similar to very recent literature \cite{venkataramanan2020attention}, we integrate supervision on attention maps during training. In particular, attention maps $\aaa \in \real^{\Omega_i}$ are generated from the latent mean vector $\zz_{\mu}$, by using Grad-CAM \cite{Selvaraju2020Grad-CAM:Localization} via backpropagation to an encoder block output $f^s_{\ttt}(\xx)$, at a given network depth $s$. Thus, for a given input image $\xx^j$ its corresponding attention map is computed as $\aaa^j= \sigma(\sum_{k}^{K} \alpha_{k} f^s_{\ttt}(\xx^j)_{k})$,where $K$ is the total number of filters of that encoder layer, $\sigma$ a sigmoid operation, and $\alpha_{k}$ are the generated gradients such that: $\alpha_{k}=\frac{1}{|\aaa^j|}\sum_{t \in \Omega_T}\frac{\partial \zz_{\mu}}{\partial \aaa^j_{k,t}}$, where $\Omega_T$ is the spatial features domain.

In \cite{venkataramanan2020attention}, authors leveraged the attention maps by enforcing them to cover the whole normal image. To achieve this, their loss function was augmented with an additional term, referred to as expansion loss, which takes the form of: $\mathcal{L}_s=\frac{1}{|\aaa|}\sum_{l \in \Omega_i}(1- \mathbf{a}^j_l)$. We can easily observe that this term resembles to an equality constraint, forcing the class activation maps to be maximum at the whole image in a pixel-wise manner (i.e., it penalizes each single pixel individually). Contrary to this work, we integrate supervision on attention maps by enforcing inequality constraints on its global target size, which allows much more flexibility, particularly when integrating the notion of expected target size, as we describe below. Indeed, as demonstrated by our results, relaxing the assumption that CAMs in normal images should cover the whole image brings a substantial performance gain. % on detecting anomalies.
Hence, we aim at minimizing the following constrained optimization problem:
\begin{align}
\label{eq:constrained_eq}
\min_{\ttt,\ppp} \quad & \mathcal L_{VAE}(\ttt,\ppp) \qquad \text{s.t.} \quad f_c(\mathbf{a}^n) \leq 0, \quad n=1,...,N  
\end{align}
   
where $f_c(\mathbf{a}^j)=(1- \frac{1}{|\Omega_i|}\sum_{l \in \Omega_i} \mathbf{a}^j_l)-p$ is the constraint over the attention map from the $j$-\textit{th} image, which enforces the generated attention map to cover the whole image, relaxed by a certain margin $p$ (in our context, $p$ defines the proportion of pixels over an entire image). To better highlight the advantages over \cite{venkataramanan2020attention}, let us take for example the case where the size proportion is $p=0.1$ (i.e., desired target size equal to 0.9, or 90\% of the image). Following the formulation in \cite{venkataramanan2020attention}, this is achieved when \textit{all the pixel predictions are equal to 0.9}, resulting in a region covering the whole image once the class activation map is thresholded. In contrast, our formulation can yield to multiple solutions, as we do not constrain individual pixels to have a 0.9 value. For instance, 90\% of the pixels having a prediction close to 1 and 10\% of the pixels with a close to 0 prediction would be a valid solution, as it satisfies the global constraint. Furthermore, note that the gradients from both terms are also different. The term in \cite{venkataramanan2020attention} leads to different gradients at each logit, while our term backpropagates the same gradient value through all the logits, based on the global target size difference. Thus, both terms are fundamentally different and lead to different solutions.% when expected target size is (0,1).}

From eq. \ref{eq:constrained_eq} we can derive an approximate unconstrained optimization problem by employing a penalty-based method, which takes the hard constraint and moves it into the loss function as a penalty term ($\mathcal{P}(\cdot)$): $\min_{\ttt,\ppp} \mathcal{L}_{VAE}(\ttt,\ppp) + \lambda \mathcal{P}(f_c(\mathbf{a}))$. Thus, each time that the constrained $f_c(\mathbf{a}^n) \leq 0$ is violated, the penalty term $\mathcal{P}(f_c(\mathbf{a}^n))$ increases.                            
\subsection{Extended log-barrier as an alternative to penalty-based functions}

Despite having demonstrated a good performance in several applications \cite{kervadec2019constrained,pathak2015constrained,he2017learning,jia2017constrained} penalty-based methods have several drawbacks. First, these unconstrained minimization problems have increasingly unfavorable structure due to ill-conditioning \cite{fiacco1990nonlinear,luenberger1973introduction}, which typically results in an exceedingly slow convergence. And second, finding the optimal penalty weight is not trivial. To address these limitations, we replace the penalty-based functions by the approximation of log-barrier\footnote{Note that this function is convex, continuous and twice-differentiable.} presented in \cite{Kervadec2019ConstrainedExtensions}, which is formally defined as:

\begin{equation}
\label{eq:log_barrier_extension}
\widetilde{\psi}_{t}(z) =
\begin{cases}
-\frac{1}{t} \log (-z) & \text{if } z \leq -\frac{1}{t^2} \\
tz - \frac{1}{t} \log (\frac{1}{t^2}) + \frac{1}{t} & \text{otherwise} ,
\end{cases}
\end{equation}

where $t$ \textit{controls} the barrier over time, % (i.e., during training), 
and $z$ is the constraint $f_c(\mathbf{a}^n)$. Thus, by taking into account the approximation in \ref{eq:log_barrier_extension}, we can solve the following unconstrained problem by using standard Gradient Descent:
\vspace{-2mm}
\begin{equation}
\label{eq:criterion}
\min_{\ttt,\ppp} \quad  \underbrace{\mathcal L_{VAE}(\ttt,\ppp)}_{\text{Standard VAE loss}} + \lambda \underbrace{\sum_{n=1}^N\widetilde{\psi}_{t}((1- \frac{1}{|\Omega_i|}\sum_{l \in \Omega_i} \mathbf{a}^n_l)-p)}_{\mathcal{L}_s:~\text{Size regularizer}}
\end{equation}

In this scenario, for a given $t$, the optimizer will try to find a solution with a good compromise between minimizing the loss of the VAE and satisfying the constraint $f_c(\mathbf{a}^n)$.

\subsection{Inference}
During inference, we use the generated attention as an anomaly saliency map. In order to avoid saturation caused by large activations, we replaced the sigmoid operation by a minimum-maximum normalization. Finally, the map is thresholded to locate the anomalous patterns in the image (different strategies to set the threshold are discussed later). Instead of inverting the generated attention maps, as in \cite{venkataramanan2020attention}, we found that anomalous regions actually produce stronger gradients, which aligns with recent observations recent observations on natural \cite{liu2020towards} and brain MRI \cite{Baur2021AutoencodersStudy} images. We believe that these larger gradients result in strongest activations, as CAM based on GradCAM are weighted by the input gradient.

\vspace{-3mm}

\section{Experiments}
\label{sec:experiments}

\subsection{Experimental setting}

\paragraph{Datasets.}The experiments described in this work were carried out using the popular BraTS 2019 dataset \cite{Menze2015TheBRATSJ, Bakas2017AdvancingFeaturesJ, Bakas2018IdentifyingChallengeJ}, which contains $335$ multi-institutional multi-modal MR scans with their corresponding Glioma segmentation masks. Following \cite{Baur2019DeepImagesJ}, from every patient, $10$ consecutive axial slices of FLAIR modality of resolution $224\times224$ pixels were extracted around the center to get a pseudo MRI volume. Then, the dataset is split into training, validation and testing groups, with $271$, $32$ and $32$ patients, respectively. Following the standard literature, during training only the slices without lesions are used as normal samples. For validation and testing, scans with less than $0.01\%$ of tumour are discarded. A summary of the dataset used is presented in Table 1 in Supplemental Materials.

\vspace{-3mm}

\paragraph{Evaluation Metrics.}
We resort to standard metrics for unsupervised brain lesion segmentation, as in \cite{Baur2021AutoencodersStudy}. Concretely, we compute the dataset-level area under precision-recall curve (AUPRC) at pixel level, as well the are under receptive-operative curve (AUROC). From the former, we obtain the operative point (OP) as threshold to generate the final segmentation masks. Then, we compute the best dataset-level DICE-score ({[}DICE{]}) and intersection-over-union ({[}IoU{]}) over these segmentation masks. Finally, we compute the average DICE over single scans. For each experiment, the metrics reported are the average of three consecutive repetitions of the training, to account for the variability of the stochastic factors involved in the process.

\vspace{-3mm}

\paragraph{Implementation Details.}The VAE architecture used in this work is based on the recently proposed framework in \cite{venkataramanan2020attention}. Concretely, the convolution layers of ResNet-18 \cite{He2016DeepRecognition} are used as the encoder, followed by a dense latent space $\zz\in \real^{32}$. For image generation, a residual decoder is used, which is symmetrical to the encoder. It is noteworthy to mention that, even though several methods have resorted to a spatial latent space \cite{Baur2019DeepImagesJ,venkataramanan2020attention}, we observed that a dense latent space provided better results, which aligns to the recent benchmark in \cite{Baur2021AutoencodersStudy}. The VAE was trained during $400$ iterations with eq. (\ref{eq:VAE_vanilla}) %without any expansion constrain 
to stabilize the convergence using $\beta = 1$. Then, the proposed regularizer was integrated (equation \ref{eq:criterion}) with $t=20$ and $\lambda=10$, applied to the Grad-CAMs obtained from the first convolutional block of the encoder. We use a batch size of $32$ images, and a learning rate of $1e{-4}$ with ADAM optimizer. The reconstruction loss, $\mathcal{L}_{R}$, in eq. (\ref{eq:VAE_vanilla}) is the binary cross-entropy, and $p$ in eq. (\ref{eq:constrained_eq}) is set empirically to $0.2$. Ablation experiments reported in the experimental section, which are performed on the validation set, empirically validate these choices. The code and trained models are publicly available  on (\url{https://github.com/cvblab/anomaly_localization_vae_gcams}).   
\vspace{-3mm}

\paragraph{Baselines.}In order to compare our approach to state-of-the-art methods, we implemented prior works and validated them on the dataset used, under the same conditions. First, we use residual-based methods to match the recently benchmark on unsupervised lesion localization in \cite{Baur2021AutoencodersStudy}. We also include recently proposed methods that integrate CAMs to locate anomalies. For both strategies, the training hyper-parameters and AE/VAE architectures were similar to the implementation of the proposed method. \textbf{\textit{Residual methods}}, given an anomalous sample, aim to use the AE/VAE to reconstruct its normal counterpart. Then, they obtain an anomaly localization map using the residual between both images such that $\mm = \lvert \xx -\hat{\xx} \rvert$, where $\lvert \cdot \rvert$ indicates the absolute value. On the AE/VAE scenario, we include methods which propose modifications over vanilla versions, including context data augmentation in Context AE \cite{zimmerer2019context}, Bayesian AEs \cite{pawlowski2018unsupervised}, Restoration VAEs \cite{Chen2020UnsupervisedPriorMedIA}, an adversial-based VAEs, AnoVAEGAN \cite{Baur2019DeepImagesJ} and a recent GAN-based approach, F-anoGAN \cite{schlegl2019f}. For methods including adversarial learning, DC-GAN \cite{Radford2016UnsupervisedNetworks} is used as discriminator. During inference, residual maps are masked using a slight-eroded brain mask, to avoid noisy reconstructions along the brain borderline. \textbf{\textit{CAMs-based}}: we use Grad-CAM VAE \cite{liu2020towards}, which obtains regular Grad-CAMs on the encoder from the latent space $\zz_{\mu}$ of a trained vanilla VAE. Concretely, we include a disentaglement variant of CAMs proposed in this work, which computes the combination of individually-calculated CAMs from each dimension in $\zz_{\mu}$, referred to as Grad-CAM$_{D}$ VAE. We also use the recent method in \cite{venkataramanan2020attention} (CAVGA), which applies a L1 penalty on the generated CAM to maximize the attention. In contrast to our model and \cite{liu2020towards}, the anomaly mask in \cite{venkataramanan2020attention} is generated by focusing on the regions not activated on the saliency map such that $\aaa = 1 - CAM$, hypothesizing that the network has learnt to focus only on normal regions. Then, $\aaa$ is thresholded with 0.5 to obtain the final anomaly mask $\mm \in \mathbb{R}^{\Omega_i}$. For both methods, the network layer to obtain the Grad-CAMs is the same as in our method. 

\vspace{-3mm}

\paragraph{How the attention maps are thresholded?}Nearly all prior approaches resort to anomalous images to define the threshold to obtain the final segmentation masks. In particular, these methods look at the AUPRC on the anomalous images, which is then used to compute the threshold value. Having access to images with anomalies is unrealistic in practice, and the value found might be biased towards the images employed. To alleviate this issue, we also evaluate our model in the scenario where the threshold is simply set to 0.5.

%\vspace{-3mm}

\subsection{Results}
\label{sec:results}

%% --------------------------------
% Soa vs Proposed

\paragraph*{Comparison to the literature.}The quantitative results obtained by the proposed model and baselines on the test cohort are presented in Table \ref{tab:test_results}. Results from the baselines range between {[}$0.056$-$0.511${]}(AUPRC) and {[}$0.188$-$0.525${]} (DICE), which are in line with previous literature \cite{Baur2021AutoencodersStudy}. We can observe that the proposed methodology (\textit{last row}) outperforms previous approaches by a large margin , with a substantial increase of $\sim$24\% and $\sim$18\% in terms of AUPRC and DICE, respectively, compared to the best prior model, i.e., F-anoGAN. \textbf{Threshold computed from normal or abnormal images?} Furthermore, we can observe that while employing images with anomalies to select the optimal threshold yields the best result, this might be unrealistic in a fully unsupervised scenario. Nevertheless, an interesting property of our approach is that \textit{it can still achieve large performance gains without having access to anomalous images to define the threshold}, unlike prior works.

\begin{table}[h]
\centering
\small
\begin{tabular}{|l|l|l|l|l|l|}
\hline
\multicolumn{1}{|c|}{\textbf{Method}} & \multicolumn{1}{c|}{\textbf{AUROC}} & \multicolumn{1}{c|}{\textbf{AUPRC}} & \multicolumn{1}{c|}{\textbf{{[}DICE{]}}} & \multicolumn{1}{c|}{\textbf{{[}IoU{]}}} & \multicolumn{1}{c|}{\textbf{DICE ($\mu\pm\sigma$)}} \\ \hline\hline
CAVGA \cite{venkataramanan2020attention}*  & $0.726$ & $0.056$ & $0.188$ & $0.104$ & $0.182\pm0.096$ \\ \hline
Bayesian VAE \cite{pawlowski2018unsupervised} & $0.922$ & $0.193$ & $0.342$ & $0.206$ & $0.329\pm0.115$ \\ \hline
AnoVAEGAN \cite{Baur2019DeepImagesJ} & $0.925$ & $0.232$ & $0.359$ & $0.221$ & $0.349\pm0.115$ \\ \hline
Bayesian AE \cite{pawlowski2018unsupervised} & $0.940$ & $0.279$ & $0.389$ & $0.242$ & $0.375\pm0.130$ \\ \hline
AE & $0.937$ & $0.261$ & $0.397$ & $0.248$ & $0.386\pm0.125$\\ \hline
%Grad-CAM VAE \cite{liu2020towards}  & $0.931$  & $0.288$ & $0.374$ & $0.231$ & $0.339\pm0.150$ \\ \hline
Grad-CAM$_{D}$ VAE \cite{liu2020towards}  & $0.941$  & $0.312$ & $0.400$ & $0.250$ & $0.361\pm0.164$ \\ \hline
Restoration VAE \cite{Chen2020UnsupervisedPriorMedIA}  & $0.934$ & $0.352$ & $0.403$ & $0.252$ & $0.345\pm0.186$ \\ \hline
Context VAE \cite{zimmerer2019context} & $0.939$ & $0.271$ & $0.406$ & $0.255$ & $0.394\pm0.126$ \\ \hline
Context AE \cite{zimmerer2019context} & $0.940$ & $0.278$ & $0.411$ & $0.259$ & $0.399\pm0.126$\\ \hline
VAE \cite{Baur2019DeepImagesJ,Zimmerer2020Abstract:Auto-encoders} & $0.940$ & $0.273$ & $0.411$ & $0.259$ & $0.399\pm0.127$ \\ \hline
F-anoGAN \cite{schlegl2019f} & $0.946$ & $0.511$ & $0.525$ & $0.369$ & $0.494\pm0.151$  \\ \hline

\rowcolor{gray!5}\textbf{Proposed*(th=0.5)} & $0.981$ & $0.753$ & $0.606$ & $0.439$ & $0.582\pm0.238$ \\ \hline
\rowcolor{gray!5}\textbf{Proposed} & $\mathbf{0.981}$ & $\mathbf{0.753}$ & $\mathbf{0.704}$ & $\mathbf{0.543}$ & $\mathbf{0.665\pm0.200}$ \\ \hline
 % \multicolumn{6}{l}{\scriptsize{*Threshold defined on the training set.}}\\
\end{tabular}
\caption{Comparison to prior literature. (*) indicates that  the class-specific threshold to locate the anomaly is set without accessing to abnormal images. Best results in bold.}
\label{tab:test_results}
\vspace{-5mm}
\end{table}

%% --------------------------------
% Image Level vs pixel-level expansion 

\paragraph*{Image vs. pixel-level constraint.} The following experiment demonstrates the benefits of imposing the constraint on the whole image %(i.e., $1-\sum(\aaa)$) 
rather than in a pixel-wise manner, %(i.e., $\frac{1}{|\aaa|}\sum(1-\aaa)$), 
as in \cite{venkataramanan2020attention}. In particular, we compare the two strategies when the constraint is enforced via a L2-penalty function, whose results are presented in Table \ref{tab:ablation4}. These results illustrate the superiority of our method, which is consistent across every $p$ value. 

\begin{table}[h!]
\centering
\begin{tabular}{|l|c|c|c|c|c|c|c|}
\hline
\multicolumn{1}{|c|}{\textbf{Regularization}} & \multicolumn{7}{c|}{\textbf{Size (proportion) term p}} \\ \hline
 & 0 & 0.05 & 0.10 & 0.15 & 0.20 & 0.25 & 0.30 \\ \hline\hline
L2 (pixel-level) & 0.489 & 0.288 & 0.275 & 0.329 & 0.288 & 0.264 & 0.201 \\ \hline
L2 (image-level) & 0.576 & 0.589 & 0.648 & 0.594 & 0.666 & 0.553 & 0.531 \\ \hline
\end{tabular}
\caption{Quantitative comparison, in terms of AUPRC, between enforcing the constraint at pixel-level (i.e., \cite{venkataramanan2020attention}) or at image-level (i.e., proposed approach) across different $p$ values.}
\label{tab:ablation4}
\end{table}

%% --------------------------------
% Penalties vs Extended log-barrier 
\vspace{-5mm}
\paragraph*{Extended log-barrier vs. penalty-based functions.}To motivate the choice of employing the extended log-barrier over standard penalty-based functions %to impose constraints 
in the constrained optimization problem in eq. (\ref{eq:constrained_eq}), we compare them in Table \ref{tab:ablation_reg}. %Note that $p$ is the proportion term in our formulation. 
First, it can be observed that across different $p$ values, imposing the constraint with the extended log-barrier consistently outperforms the L$_2$-penalty, with substantial performance gains. Furthermore, we empirically observe that despite any analyzed value of $p$ outperforms current sota, setting $p=0.2$ brings the largest performance gain.

\begin{table}[h!]
\centering
\begin{tabular}{|l|c|c|c|c|c|c|c|}
\hline
\multicolumn{1}{|c|}{\textbf{Regularization}} & \multicolumn{7}{c|}{\textbf{Size (proportion) term p}} \\ \hline
 & 0 & 0.05 & 0.10 & 0.15 & 0.20 & 0.25 & 0.30 \\ \hline\hline
%L1 (Penalty) & \begin{tabular}[c]{@{}c@{}}0.599\\ 0.594\end{tabular} & \begin{tabular}[c]{@{}c@{}}0.630\\ 0.619\end{tabular} & \begin{tabular}[c]{@{}c@{}}0.649\\ 0.626\end{tabular} & \begin{tabular}[c]{@{}c@{}}0.602\\ 0.610\end{tabular} & \begin{tabular}[c]{@{}c@{}}0.625\\ 0.613\end{tabular} & \begin{tabular}[c]{@{}c@{}}0.656\\ 0.645\end{tabular} & \begin{tabular}[c]{@{}c@{}}0.537\\ 0.560\end{tabular} \\ \hline
L2 (Penalty) & \begin{tabular}[c]{@{}c@{}}0.576\\ 0.581\end{tabular} & \begin{tabular}[c]{@{}c@{}}0.589\\ 0.581\end{tabular} & \begin{tabular}[c]{@{}c@{}}0.648\\ 0.631\end{tabular} & \begin{tabular}[c]{@{}c@{}}0.594\\ 0.589\end{tabular} & \begin{tabular}[c]{@{}c@{}}0.666\\ 0.632\end{tabular} & \begin{tabular}[c]{@{}c@{}}0.553\\ 0.557\end{tabular} & \begin{tabular}[c]{@{}c@{}}0.531\\ 0.521\end{tabular} \\ \hline
Extended Log Barrier \cite{Kervadec2019ConstrainedExtensions} & \begin{tabular}[c]{@{}c@{}}0.682\\ 0.653\end{tabular} & \begin{tabular}[c]{@{}c@{}}0.664\\ 0.638\end{tabular} & \begin{tabular}[c]{@{}c@{}}0.646\\ 0.625\end{tabular} & \begin{tabular}[c]{@{}c@{}}0.641\\ 0.623\end{tabular} & \textbf{\begin{tabular}[c]{@{}c@{}}0.710\\ 0.661\end{tabular}} & \begin{tabular}[c]{@{}c@{}}0.640\\ 0.625\end{tabular} & \begin{tabular}[c]{@{}c@{}}0.610\\ 0.600\end{tabular} \\ \hline
\end{tabular}
\caption{Impact of the type of regularization and the proportion term $p$ on our method. We report the AUPRC (\textit{top row}) and the best {[}DICE{]} (\textit{bottom row}).
Best results in bold.}
\label{tab:ablation_reg}
\end{table}

%% --------------------------------
% Reference to supplemental materials for more experiments

\noindent In the Supplemental Materials, we provide comprehensive ablation experiments to validate several elements of our model, and motivate the choice of the values employed in our formulation, as well as our experimental setting.

\begin{figure}[h!]
\begin{center}
\centering
\includegraphics[width=0.9\linewidth]{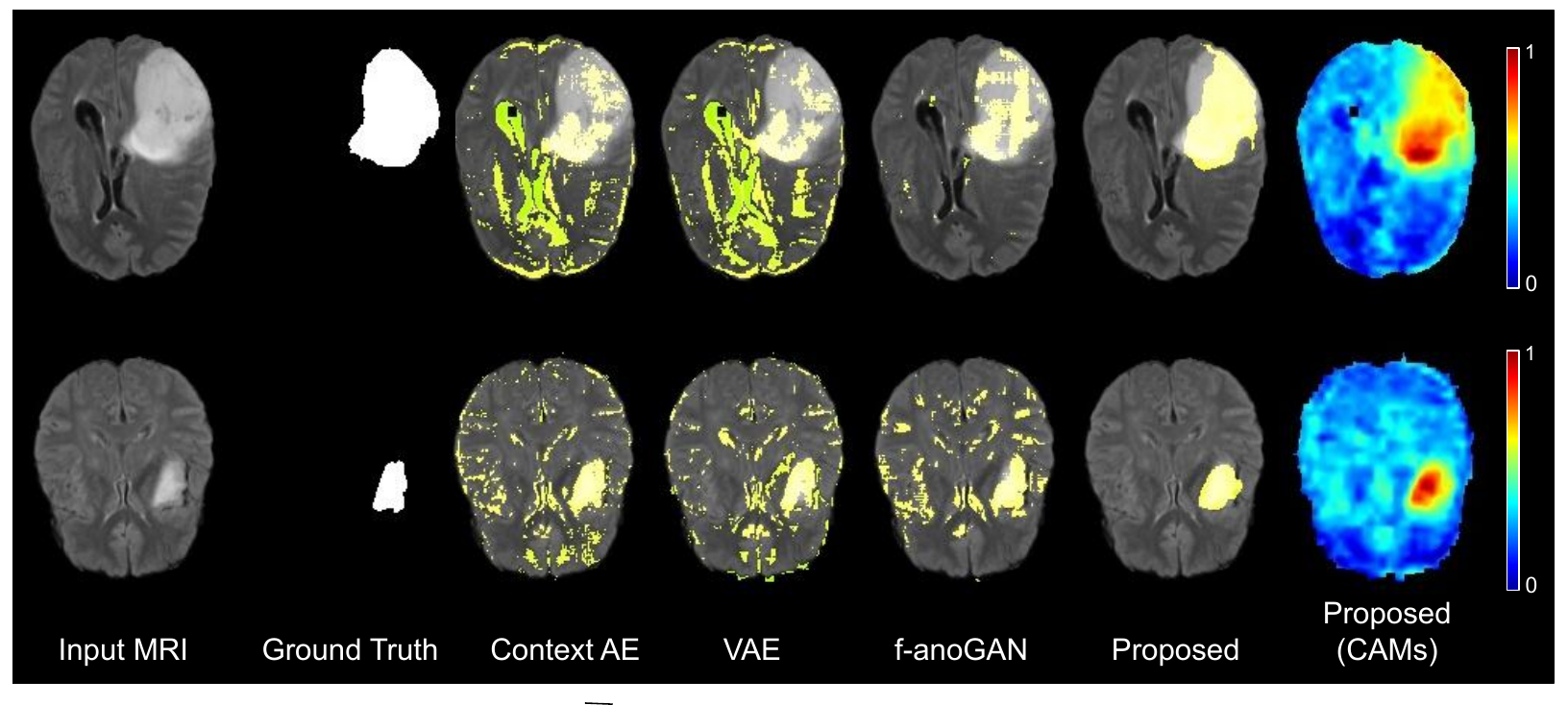} 
\caption{Qualitative evaluation of the proposed and existing high-performing methods.}
\label{fig:qualitative}
\end{center}
\end{figure}

%% --------------------------------
% Qualitative Evaluation
\vspace{-2mm}
\paragraph{Qualitative evaluation.}Visual results of the proposed and existing methods are depicted in Figure \ref{fig:qualitative}. We can observe that our approach identifies as anomalous more complete regions of the lesions, whereas existing methods are prone to produce a significant amount of false positives (\textit{top and bottom rows}) and fail to discover many abnormal pixels (\textit{top row}). 

%% --------------------------------
% Conclusions

\vspace{-1mm}
\section{Conclusions}
\label{sec:conclusions}

We proposed a novel constrained formulation for the task of unsupervised segmentation of brain lesions. In particular, we resort to generated CAMs to identify anomalous regions, which contrasts with most existing works that rely on the pixel-wise reconstruction error. Our formulation integrates a size-constrained loss that enforces the CAMs to cover the whole image in normal images. In contrast to very recent work, we tackle this problem by imposing inequality constraints on the whole target CAMs, which allows more flexibility than equality constraints over each single pixel. Last, and to overcome the limitations of penalty-based methods, we resort to an extension of standard log-barrier methods. Quantitative and qualitative results demonstrate that our model significantly outperforms prior literature on unsupervised lesion segmentation, without the need of accessing to anomalous images.    

%% --------------------------------
% Bibliography

\bibliography{egbib, references_cvblab}

\newpage
\setcounter{section}{0}
\setcounter{table}{0}

%% --------------------------------
% Supplemental material
\bigskip

\section*{Supplemental material}

\bigskip

%\vspace{-10mm}
\section{Additional dataset details}

A summary of the used dataset, with the corresponding training, validation and testing splits, after the pre-processing detailed in Section 4.1, is presented in Table \ref{dataset}.

\begin{table}[h!]
\centering
\begin{tabular}{|l|c|c|}
\hline
\multicolumn{1}{|c|}{\textbf{Partition}} & \textbf{Cases} & \textbf{Training Images} \\ \hline\hline
Training & $271$ & $268$ \\ \hline
Validation & $26$ & $-$ \\ \hline
Testing & $25$ & $-$ \\ \hline
\end{tabular}
\caption{Dataset, partition and training images used.}
\label{dataset}
\end{table}

\section{Additional ablation studies}

%% --------------------------------
% SM2 Proposed method hyper-parameters

\paragraph*{Model hyperparameters.} To better understand the behaviour of the attention constrains in the proposed model, we resort to extensive ablation experiments to determine the optimal values of several model hyperparameters: the log-barrier $t$ term, the size term $p$, the weights of the attention loss on the training, $\lambda$ and, finally, the network depth used to compute the CAMs. Firstly, we empirically fix $\lambda=10$ and use the first convolutional block output to compute CAMs, to evaluate the impact of our model with $p$ values included in $\{0, 0.05, 0.10, 0.15, 0.20, 0.25, 0.30\}$ and $t$ values in $\{10, 15, 20, 25, 50\}$. These results are reported in Table \ref{ablation_t_p}. Please note that all the results reported on the ablation studies are obtained on the validation set.

% Table on t and p hyperparameters
\begin{table}[h!]
\centering
\begin{tabular}{|l|c|c|c|c|c|c|c|}
\hline
\multicolumn{1}{|c|}{\textbf{t}} & \multicolumn{7}{c|}{\textbf{Size (proportion) term p}}\\ \hline
\multicolumn{1}{|c|}{}           & 0     & 0.05  & 0.10  & 0.15  & 0.20           & 0.25   & 0.30  \\ \hline\hline
10                               & 0.614 & 0.408 & 0.662 & 0.504 & 0.601          & 0.623  & 0.500 \\ \hline
15                               & 0.575 & 0.546 & 0.498 & 0.614 & 0.638          & 0.599  & 0.641 \\ \hline
20                               & 0.682 & 0.664 & 0.646 & 0.641 & \textbf{0.710} & 0.640 & 0.610 \\ \hline
25                               & 0.536 & 0.606 & 0.575 & 0.545 & 0.679          & 0.671  & 0.680 \\ \hline
50                               & 0.476 & 0.606 & 0.636 & 0.685 & 0.539          & 0.657  & 0.607 \\ \hline
\end{tabular}
\caption{Ablation study on the impact of $p$ and $t$ in the proposed formulation, where dataset specific AUPRC results are presented. Bold highlights the best performing configuration.}
\label{ablation_t_p}
\end{table}

We now validate the level depth from the encoder used to obtain the CAMs (i.e., network depth $s$ in Section 3.2), with the best configuration from the previous ablation in Table \ref{ablation_t_p}. Results are presented in Table \ref{ablation_cams}, from which we can observe that %the earlier the attention mechanism is maximized, the better the results are. 
maximizing the attention in early layers leads to better results than in deeper layers. This could be produced by the better spatial definition of early layers, and the benefits that the proposed constrain produces in its later layers, which receive information from the whole image.

\begin{table}[h!]
\centering
\begin{tabular}{|l|c|c|c|c|}
\hline
            & Conv1          & Conv2 & Conv3 & Conv4            \\ \hline\hline
AUPRC       & \textbf{0.710} & 0.621 & 0.456 & 0.274            \\ \hline
{[}DICE{]}  & \textbf{0.661} & 0.454 & 0.292 & 0.276            \\ \hline
\end{tabular}
\caption{Ablation study on network depth to compute CAMs. Dataset specific AUPRC is presented for each possible configuration. Best performance highlighted in bold.}
\label{ablation_cams}
\end{table}

Next, in Table \ref{ablation_delta} we study the optimal weight to balance the proposed attention loss, by evaluating the performance of our formulation across several $\lambda$ values. %using the following range of values: $\lambda=\{0.01, 0.1, 1, 10\}$. 
The experiments presented on the main paper are obtained using the best configuration: $t=20$, $p=0.20$, $\lambda=10$, with CAMs being obtained form the first convolutional block.

% Table on delta values Level
\begin{table}[h!]
\centering
\begin{tabular}{|l|c|c|c|c|c|}
\hline
           & \multicolumn{5}{c|}{$\lambda$} \\ \hline
           & 0.01   & 0.1   & 1  & 10        & 100  \\ \hline\hline
AUPRC      & 0.150  & 0.443 & 0.609   & \textbf{0.710} & 0.587      \\ \hline
{[}DICE{]} & 0.207  & 0.502 & 0.609   &\textbf{0.661} &  0.587    \\ \hline
\end{tabular}
\caption{Ablation study on the influence of attention expansion losses in relation to its relative weight, $\lambda$. Dataset specific AUPRC and DICE are presented for each validated value. Bold numbers indicate the best performance.}
\label{ablation_delta}
\end{table}

% Table on CAMs Level

\paragraph*{Number of slices to generate the pseudo-volumes.} In our experiments, we followed the standard literature \textcolor{red}{[8]} to generate the pseudo-labels for validation and testing. Nevertheless, it is unclear in unsupervised anomaly detection of brain lesions the appropriate number of slices used from the MRI scans. We now explore the impact of including more slices in these pseudo-volumes, which increase the variability of normal samples. In this line, we hypotetize that the dimension of the VAE latent space may be a determining factor in absorbing this increased variability. The appropriate $\zz$ dimension is also unclear in the literature. For instance, \textcolor{red}{[8]} uses $\zz=128$, while \textcolor{red}{[6]} uses $\zz=64$, and we obtained better results using $\zz=32$. To validate the proposed experimental setting and latent space dimension, we now present results using increasing number of slices around the axial midline $N=\{10, 20, 40\}$, and two different latent space dimensions $\zz=\{32, 128\}$ for both a standard VAE and our proposed model, in Table \ref{ablation_N_images}.  We can observe that despite the gap between the two method is reduced as the number of slides is increased, this difference is still significant. Finally, we can observe that an increasing on $\zz$ dimension does not produce gains in performance in any case. Note that the model hyperparameters used are optimized for $z=32$, and $N=10$, which also could produce some underestimation of the proposed model performance when $N$ increases.

\begin{table}[h]
\centering
\begin{tabular}{|l|c|c|c|c|}
\hline
\multicolumn{1}{|c|}{\textbf{Method}} & \textbf{zdim} & \multicolumn{3}{c|}{\textbf{N slices}} \\ \hline
    &  & 10 & 20 & 40 \\ \hline\hline
    
\multicolumn{1}{|c|}{\multirow{2}{*}{Proposed}} & 32 & \begin{tabular}[c]{@{}c@{}}0.710\\ 0.661\end{tabular} & \begin{tabular}[c]{@{}c@{}}0.581\\ 0.577\end{tabular} & \begin{tabular}[c]{@{}c@{}}0.572\\ 0.576\end{tabular} \\ \cline{2-5} 
\multicolumn{1}{|c|}{} & 128 & \begin{tabular}[c]{@{}c@{}}0.601\\ 0.599\end{tabular} & \begin{tabular}[c]{@{}c@{}}0.554\\ 0.562\end{tabular} & \begin{tabular}[c]{@{}c@{}}0.559\\ 0.556\end{tabular} \\ \hline

\multirow{2}{*}{VAE} & 32 & \begin{tabular}[c]{@{}c@{}}0.275\\ 0.398\end{tabular} & \begin{tabular}[c]{@{}c@{}}0.259\\ 0.373\end{tabular} & \begin{tabular}[c]{@{}c@{}}0.229\\ 0.362\end{tabular} \\ \cline{2-5} 
    & 128 & \begin{tabular}[c]{@{}c@{}}0.252\\ 0.368\end{tabular} & \begin{tabular}[c]{@{}c@{}}0.250\\ 0.384\end{tabular} & \begin{tabular}[c]{@{}c@{}}0.217\\ 0.347\end{tabular}\\ \hline
\end{tabular}
\caption{Ablation study on number of axial slices around the center used from MR brain volumes, and lattent space dimension, for the proposed model and an standard VAE.  We report the AUPRC (\textit{top row}) and the best {[}DICE{]} (\textit{bottom row}).}
\label{ablation_N_images}
\end{table}

\paragraph*{On the impact of the reconstruction losses.}We evaluate the effect of including several well-known reconstruction losses in our formulation: SSIM [38] and L$_2$-norm. Table \ref{table:recons_loss} reports the results from these experiments, where we can observe that, while BCE and SSIM reconstruction losses yield the best performances, integrating the L$_2$-norm loss in our formulation degrades the performance of the proposed model.    

\begin{table}[h!]
\centering
\small
\begin{tabular}{|l|c|c|c|}
\hline
       & \textbf{BCE} & \textbf{L2 norm} & \textbf{SSIM} \\ \hline\hline
AUPRC  & $\mathbf{0.710}$ & $0.600$ & $0.679$ \\ \hline
[DICE] & $\mathbf{0.661}$ & $0.612$ & $0.649$ \\ \hline
\end{tabular}
\caption{Ablation study on the reconstruction losses for the proposed approach. Best results in bold.}
\label{table:recons_loss}
\end{table}

%% --------------------------------
% SM3 Using normal domain statistics for threshold 

\paragraph*{Using statistics from normal domain for anomaly localization threshold} As mentioned along the manuscript, a main limitation of unsupervised anomaly localization methods is the need of using anomalous images to set a threshold on the obtained heatmaps to locate anomalies. Several methods \textcolor{red}{[6]} have discussed the possibility of using a given percentile from the normal images (i.e., no anomalies) distribution to set the threshold. An ablation study on the percentile value is presented in Table \ref{ablationpercentille} for our proposed model and the best performing baseline. Compared to the best baseline method in Table 1 of the main manuscript, i.e., F-anoGAN, our model substantially yields better performance. Nevertheless, we found that the best results are obtained on the percentile $95\%$, whereas \textcolor{red}{[6]} found the operative performance on the percentile $98\%$. This suggests that, even though not used directly, anomalous images are still required to find the optimal value.

%Results show that still, anomalous images are required to find the optimum value. For instance, \cite{Baur2019DeepImages} found the operative performance on the percentile $98\%$, while we obtain better results with $95\%$. As already indicated in Section \ref{sec:results}, our proposed method can achieve large performance gains compared to previous literature by simply fixing the anomaly threshold to $0.5$.

\begin{table}[h!]
\centering
\begin{tabular}{|l|c|c|c|c|c|c|}
\hline
         & OP & th=0.5 & p85   & p90    & p95 & p98 \\ \hline\hline
Proposed & $0.661$ & $0.579$ &  $0.498$ & $0.602$ & $0.657$ & $0.597$ \\ \hline
%VAE      & $0.383$  & $-$ &  $0.252$ & $0.293$ & $0.259$ & $0.378$ \\ \hline
F-anoGAN & $0.525$  & $-$ &  $0.310$ & $0.390$ & $0.505$ & $0.488$ \\ \hline
\end{tabular}
\caption{Ablation study on threshold values from normal images. p$X$ indicates the average percentile used on the training set (normal images) to compute the segmentation threshold. OP indicates the operative point from area under precision-recall curve, using all validation dataset, which contains anomalous images. The metric presented is the dataset-level DICE.}
\label{ablationpercentille}
\end{table}

%% --------------------------------
% SM4 Model parameters 

\paragraph*{Model parameters.} In this section, we compare our formulation to existing approaches in terms of model complexity. %pay attention to another interesting property of CAM-based anomaly localization. 
Since previous residual-based methods require the generation of normal counterparts from anomalous images, they typically integrate an additional discriminator to create more realistic images, and require to use the trained generative decoder during inference. On the other hand, our proposed formulation only requires the encoder part of the network to localize anomalies, which reduces the number of required parameters, as indicated in Table \ref{parameters}. On the other hand, as highlighted in previous works \textcolor{red}{[19]} the cost of adding a single constraint is negligible. 

\begin{table}[h!]
\centering
\begin{tabular}{|l|c|c|}
\hline
\multicolumn{1}{|c|}{\textbf{Method}} & \multicolumn{2}{c|}{\textbf{$\sim$Parameters (millions)}}                     \\ \hline
                                      & \multicolumn{1}{c|}{Train} & \multicolumn{1}{c|}{Inference}  \\ \hline\hline
Context VAE \textcolor{red}{[39]}                            & 15.0     & 15.0    \\ \hline
VAE \textcolor{red}{[6, 40]} & 15.0     & 15.0    \\ \hline
F-anoGAN \textcolor{red}{[33]}                                      & 17.8     & 15.0    \\ \hline
Proposed                                                          & 15.0     & 13.3    \\ \hline
\end{tabular}
\caption{Parameters of the proposed method and best performing baselines during both, training and inference stages.}
\label{parameters}
\end{table}

%% --------------------------------
% SM5 Qualitative results 

\paragraph*{Additional qualitative results.}

In the following Figure \ref{fig:qualitative_sup}, we show complementary examples of the proposed method performance.

\begin{figure}[h!]
\begin{center}
\centering
\includegraphics[width=0.9\linewidth]{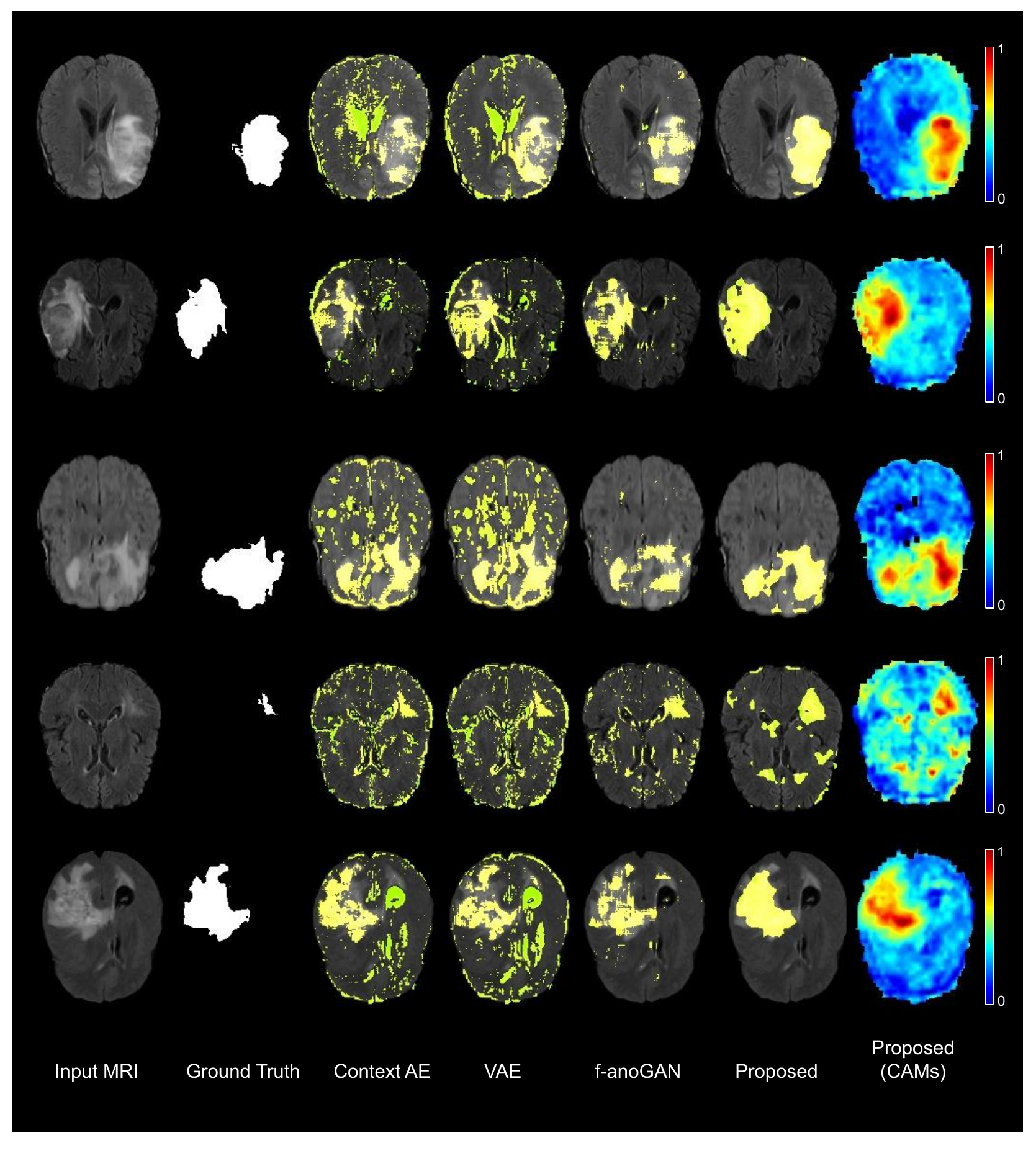} 
\end{center}
\end{figure}

\makeatletter
\setlength{\@fptop}{0pt}
\makeatother

\begin{figure}[h!]
\begin{center}
\centering
\includegraphics[width=0.9\linewidth]{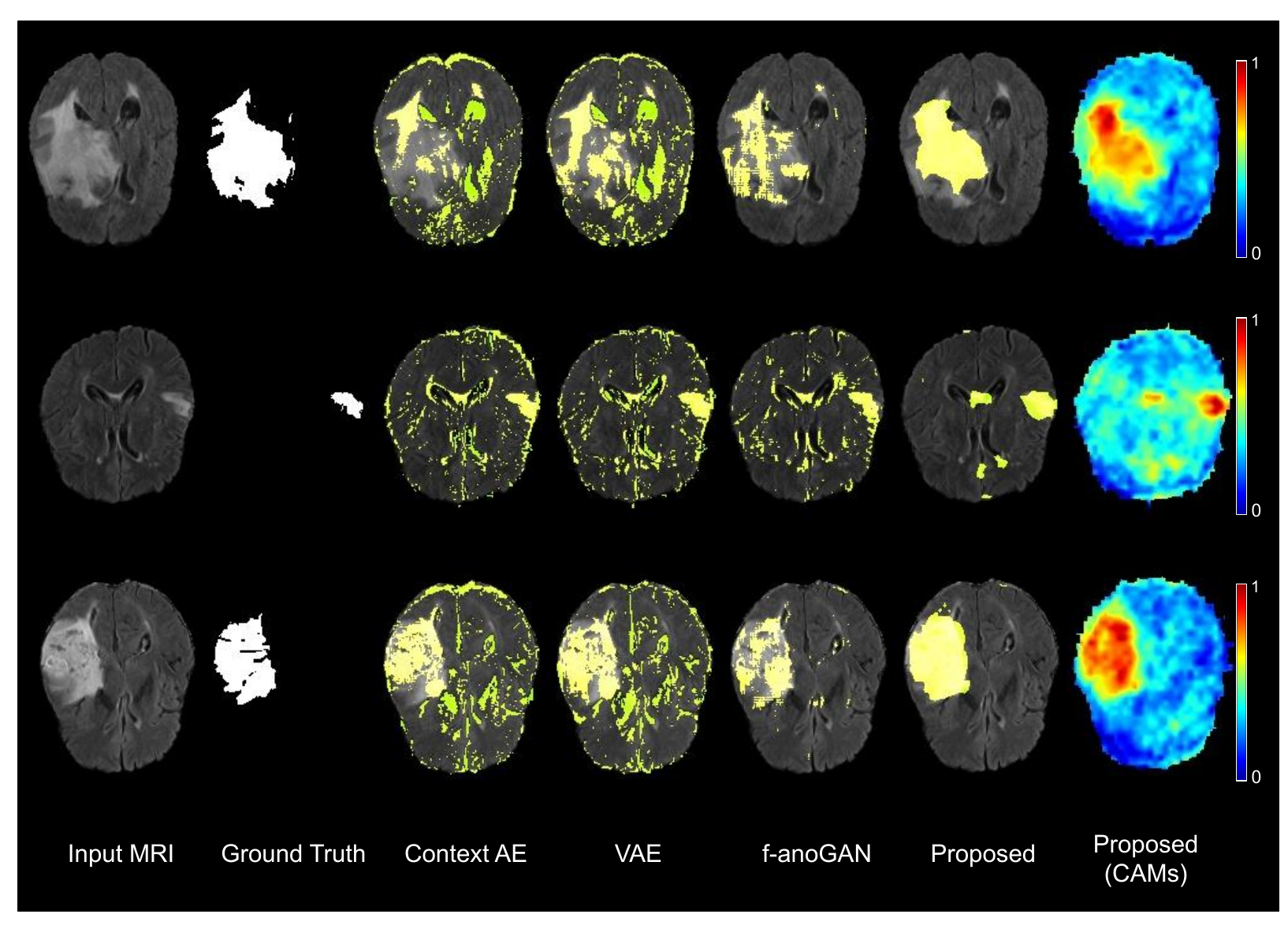} 
\caption{Qualitative evaluation of our method.}
\label{fig:qualitative_sup}
\end{center}
\end{figure}

\begin{comment}

\begin{figure}[t!]
\begin{center}
\centering
\includegraphics[width=0.9\linewidth]{images/BMVC_qualitative_2.pdf} 
\caption{Qualitative evaluation of our method.}
\label{fig:qualitative_sup}
\end{center}
\end{figure}

\end{comment}

%% --------------------------------
% Bibliography

%\bibliography{egbib, references_cvblab_downloaded}

\end{document}